\documentclass{iopart}

\usepackage{graphicx}

\begin{document}

\title{Directional Quasi-Phase Matching in Curved Waveguides}

\author{Rolf T. Horn$^1$, Gregor Weihs$^{1,2}$}
\address{$^1$Institute for Quantum Computing and Department of Physics and Astronomy,
University of Waterloo, 200 University Avenue W, Waterloo, Ontario, N2L 3G1, Canada.\ead{rhorn@iqc.ca}}
\address{$^2$Institut f\"ur Experimentalphysik, Universit\"at Innsbruck,
Technikerstra{\ss}e 25, 6020 Innsbruck, Austria. \ead{gregor.weihs@uibk.ac.at}}

\begin{abstract}
In materials that do not allow birefringent phase-matching or periodic poling we propose to use waveguides to 
exploit the tensor structure of the second order nonlinearity for quasi-phase matching of nonlinear interactions. 
In particular, we concentrate on curved waveguides in which the interplay between the propagation direction, 
electric field polarizations and the nonlinearity can change the strength and sign of the nonlinear interaction 
periodically to achieve quasi-phase matching.
\end{abstract}

Nonlinear optics enables many technologies in classical and quantum communication. The best known application is 
second-harmonic generation (SHG) \cite{Kleinman66a}, the frequency doubling of an electromagnetic wave. Quantum 
mechanically the reverse process can happen as well and a single photon can split into two photons under energy 
conservation. This process is called spontaneous parametric down-conversion (SPDC) and has been successfully used 
for generating entangled photon pair states \cite{Kwiat95b}, or heralded single photons 
\cite{Bocquillon09,Razavi09}. SPDC is the workhorse for quantum communication \cite{Bouwmeester97a,Erven08a} and 
optical quantum information processing \cite{Pan03b}.

The second order term in the perturbative expansion of the electric polarization caused by a driving field is 
responsible for SHG, sum- and difference frequency generation (SFG and DFG), and SPDC. In all of these processes 
waves with vastly different frequencies interact. Dispersion causes the generated waves to de-synchronize from 
the driving field.  This is detrimental to the growth of these new frequencies, and one has to take special 
measures to avoid this and achieve phase-matching. In birefringent materials one can phase-match by choosing the 
polarizations, crystal cut, and crystal temperature so that the birefringence effectively cancels the dispersion. 
Furthermore, by structuring the material into waveguides with suitable modes \cite{Scaccabarozzi06a}, by 
introducing form birefringence \cite{Leo99,Helmy06a} one can achieve matching effective refractive indices.

All other techniques employ some form of quasi-phase matching (QPM) by judiciously introducing periodicity to  
rephase the waves \cite{Armstrong62}. Photonic crystals offer the highest flexibility by introducing periodicity 
both in the linear and nonlinear properties of the material \cite{Weihs06}. Traditional QPM methods only 
periodically reverse the sign of the nonlinearity. The most common implementation is by periodic poling of a 
ferroelectric. The period is chosen so as to reverse the sign just when the accumulated phase difference begins 
to drive energy back into the exciting field. The sign reversal effectively resets the phase and allows the 
conversion to continue. Waveguides have the added benefit that the light stays contained and other optical 
elements can be integrated into the structure. In this article, we introduce another kind of quasi-phase-matching 
that employs waveguides. We call this method directional quasi-phase-matching (DQPM).

DQPM makes use of the fact that the strength of the nonlinear interaction is governed by a tensor 
$\chi^{(2)}_{ijk}$ -- or $d_{ij}$ in the usual contracted notation \cite{Boyd08}. Therefore, the coupling 
strength between three waves depends on the orientation of the three fields with respect to the crystal and each 
other. If we assume that the field propagation direction in a waveguide adiabatically follows the direction of 
that waveguide and the field polarizations stay approximately transverse to the propagation direction, then the 
coupling can be modulated by changing the waveguide direction. This modulation can affect both the sign and 
magnitude of the coupling. To our knowledge, apart from some earlier work on bulk materials \cite{Haidar04} this 
effect was only noted very recently in connection with nonlinear optics in disk and ring resonators 
\cite{Dumeige06a,Yang07a,Kuo09}.

In what follows, we demonstrate the conditions for DQPM as they apply to SHG and the related phenomena in curved 
waveguides. We discuss various curve shapes in relation to the expected amplification of SHG, their bend loss and 
modal behavior. Our analysis also applies to other material systems -- in particular other compounds that 
crystallize in the zincblende structure -- but presently we focus on the $\mathrm{Al_xGa_{1-x}As}$ material 
system. $\mathrm{Al_xGa_{1-x}As}$ is a common optoelectronic material that is transparent for telecommunications 
frequencies and has a high optical nonlinearity (100~pm/V). Additionally, semiconductor lasers and other active 
components are manufactured in this system, making it an ideal choice for further integration.

In our analysis, we begin with the standard treatment of SHG in straight waveguides and modify the equations to 
account for directional changes. We neglect losses and use the slowly varying amplitude approximation. We further 
assume that the fields are mainly transverse to the direction of propagation. For bulk $\mathrm{Al_xGa_{1-x}As}$ 
which belongs to the $\bar 43m$ point group, the only non-zero elements of the second order nonlinear tensor are 
$d_{14}$=$d_{25}$=$d_{36}$, where, as usual the labels are $1,2,3 = x,y,z$ and $1,2,3,4,5,6 = xx,yy,zz,yz,xz,xy$ 
in the first and second indices, respectively \cite{Boyd08}. This means that the two fundamental field modes and 
the second harmonic are only coupled if there are nonzero field amplitudes along every one of the 
crystallographic axes $x$, $y$, and $z$.\textsuperscript{\footnotemark}

\footnotetext{In tightly confining Bragg reflection waveguide \cite{Helmy06a} samples provided by A. Helmy we 
have recently observed experimentally that other interactions (e.g. all polarizations equal) are possible that 
wouldn't be allowed in the bulk material. Such interactions may occur because the symmetry is broken by the layer 
structure of the respective waveguides.}

\begin{figure}
  \centerline{\includegraphics[width=3cm]{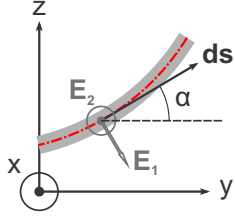}}
  \caption{Crystal coordinates, waveguide tangent vector $\mathbf { ds}$ and its angle $\alpha$ with the $y$ 
axis.}
  \label{fig:coords}
\end{figure}

To develop a visual sense of DQPM, consider two co-ordinate systems.  One is static and oriented with respect to 
the cubic symmetry of the AlGaAs lattice, the other co-moves with the direction of propagation of light in the 
waveguide. For $\mathrm{Al_xGa_{1-x}As}$, epitaxial growth is typically done on (100) surfaces ($y-z$ plane), 
therefore a waveguide is confined to lie in this plane. We denote the waveguide propagation direction tangent 
vector as $\mathbf{ds}$ as shown in figure~\ref{fig:coords}. If $\mathbf{ ds}$ is parallel to $z$, then all 
guided fields will normally be polarized along $y$ or $x$. In this case the effective nonlinearity will be zero, 
and there won't be any nonlinear conversion, because none of the fields have a $z$ component. On the other hand, 
if the waveguide is directed at $\pm 45^\circ$ to the $y$-axis, say $\mathbf{ds} \propto \mathbf{\hat 
y}+\mathbf{\hat z}$, then its two transverse polarizations are parallel to $\mathbf{\hat x}$ and parallel to 
$\mathbf{\hat y}+\mathbf{\hat z}$. In this situation the nonlinear process can, at least in principle, occur.

We will now use the dynamical equations for the amplitudes of plane waves in second harmonic generation, assuming 
that these amplitudes vary slowly compared to the wavelength. The plane wave approach is justified as long as the 
waveguide modes that are involved are mostly transverse, however, the coupling strengths will be off because the 
transverse mode structure is being neglected here. If $s$ is the arc-length parameter of the waveguide curve, the 
change in the amplitude of the second harmonic field $E_2(s)$ is given by
\begin{equation}
    \frac{d}{ds}E_2(s)=-i\frac{\omega_2}{n_2 c}\frac{d_\mathrm{\scriptsize eff}}{2}(E_1)^2e^{i\Delta{k}s}
\end{equation}
The subscript indices 1 and 2 represent the fundamental and second harmonic fields respectively. These fields are 
defined by their respective amplitudes ($E$), frequencies ($\omega$), and wavevectors ($k$), and the phase 
mismatch $\Delta{k}=k_2-2k_1>0$. $n_2$ is the effective refractive index for the second harmonic field.

We consider the generation of vertically polarized ($x$) second harmonic light ($E_2$) by injecting horizontally 
(in the $y-z$ plane) polarized fundamental light ($E_1$) into the guide. Therefore we evaluate the effective 
nonlinearity $d_\mathrm{\scriptsize eff}$ as
\begin{eqnarray}
    d_\mathrm{\scriptsize eff}&=&\sum_{ijk}d_{ijk}a_{2i}a_{1j}a_{1k}\\ &=&d_{14}a_{2x}a_{1y}a_{1z} =
d_{14}a_{1y}a_{1z},
\end{eqnarray}
where $a_{1}$ and $a_{2}$ are unit vectors in the polarization directions of the fields $E_1$ and $E_2$ 
respectively. Further $a_{2x} = 1$ because the polarization of the second harmonic field is always along $x$.

We now consider arbitrarily directed waveguide structures and reinterpret $s$ as the effective path length along 
the waveguide. In the simplest approximation $s$ will be measured along the midline of the guide. For a more 
realistic treatment one would use the corrected propagation constants for the modes of a bent waveguide, or short 
of a full two- or three-dimensional simulation, use the path taken by the field maximum. We retain the adiabatic 
slowly varying amplitude approximation for waves that could either be plane waves or confined modes with 
quasi-transverse polarizations. Also we assume that the magnitude of the phase mismatch between the fundamental 
and second harmonic waves remains constant ($\Delta{k}\neq\Delta{k(s)})$.

We define the local direction of the waveguide by its angle $\alpha(s)$ with the $y$-axis, resulting in an 
expression for the in-plane polarization vector $\mathbf a_1$ that is given by
\begin{equation}
    \mathbf a_1(s)=\left(\begin{array}{c} 0  \\ -\sin\alpha(s) \\ \cos
\alpha(s)\end{array} \right).
\end{equation}

Making the substitutions for $d$ we get
\begin{equation}
    \frac{\mathrm d E_2}{\mathrm d s} = i \frac{\omega_2 d_{14}}{n_2 c} \sin[2\alpha(s)]
E_1^2 e^{i(\Delta{k}) s}
    \label{eq:shgde}
\end{equation}

The growth of $E_2$ is now clearly dependent on both the distance traveled and the orientation of the waveguide, 
$\alpha(s)$. It is zero for a waveguide whose propagation direction is along any of the crystal axes and can have 
positive or negative sign.

We will assume that the exciting field $E_1$ has constant amplitude, i.e. is not depleted, and that $E_2(s=0)=0$. 
For straight guides with $\sin 2\alpha(s)=\mbox{const.}\geq0$ and $\Delta{k(s)}\neq 0$, (\ref{eq:shgde}) 
describes the familiar behavior for non-phase-matched SHG. That is, $E_2$ oscillates with a period of 2$L_c$ 
where $L_c=\pi/\Delta k$, between zero and it's maximum value along the guide.

\begin{figure}
    \centerline{\includegraphics[width=4.3cm]{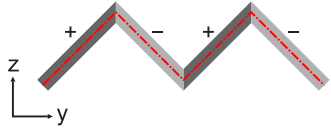}}
    \caption{First order directional quasi-phase matching.  In this highly idealized structure we assume no phase 
shifts, loss, or out of plane field rotations occurring at the sharp $90^\circ$ bends. The change in second 
harmonic amplitude (right side of (\ref{eq:shgde})) switches from positive to negative because of the 
$\sin(2\alpha(s))$ term, depending on the quadrant of $\alpha$. For first order QPM each segment have to be 
exactly $L_c$ long, so that the sign change occurs precisely when, due to the dispersion, the phase fronts 
between the pump and SHG fields have acquired a $\pi$ phase between each other.}
    \label{fig:1orderdqpm}
\end{figure}

DQPM can thus mimic the behavior of the usual periodic poling of a ferroelectric nonlinear crystal. In a 
``zig-zag'' shaped guide with segments of length $L_c$ that alternate between the $\alpha=\pi/4$ and 
$\alpha=-\pi/4$ directions the nonlinear polarization switches its sign as in periodic poling. 
Figure~\ref{fig:1orderdqpm} shows the layout of the structure on the wafer.

Clearly such sharp bends would be difficult, if not impossible to make, given that $L_c$ for the doubling of 
light at 1600~nm wavelength to 800~nm in bulk $\mathrm{Al_{0.4}Ga_{0.6}As}$ is just under two microns. Instead of 
bends it would be better to use total internal reflection mirrors, etched into the sample \cite{Kim06}, with an 
appropriate treatment of the phase change upon reflection. In any case, the losses would likely be very high, 
therefore we will now consider perturbations of first order DQPM as well as higher order phase matching to allow 
for milder bending. The fact that one has more control over the effective nonlinearity in DQPM than in poling 
reflects in the fact that even order QPM is possible here. As an example, we will look at a second order QPM 
scheme, with the main advantage that it only requires $45^\circ$ bends.

\begin{figure}
    \centerline{\includegraphics[width=3cm]{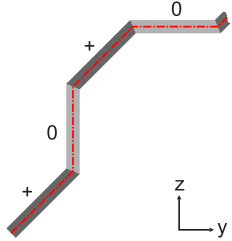}}
    \caption{This structure is an example of an even order quasi phase matching scheme -- something that is 
unavailable with periodic poling methods. This is because poling cannot turn off the non-linearity, only 
reverse it.  It also allows somewhat milder bending. However, similar growth of SHG to the first order scheme 
will require twice the distance.}
    \label{fig:2orderdqpm}
\end{figure}

\begin{figure}
    \centerline{\includegraphics[width=7cm]{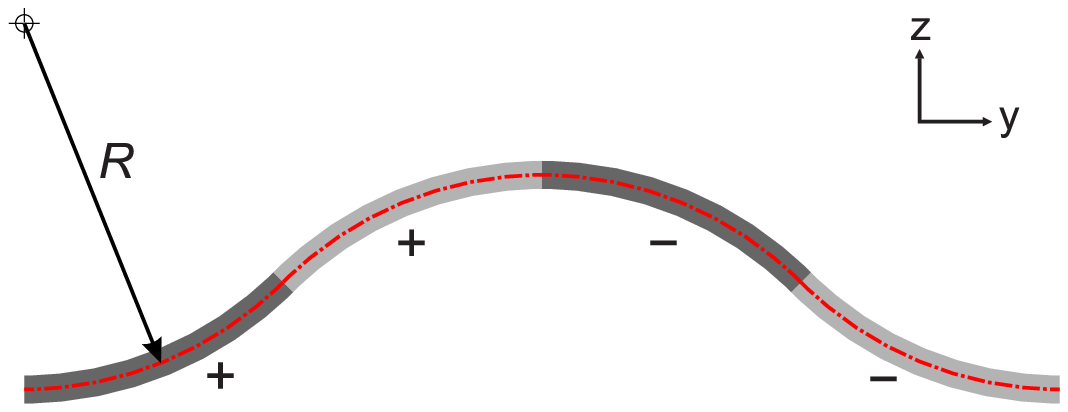}}
    \centerline{\includegraphics[width=6cm]{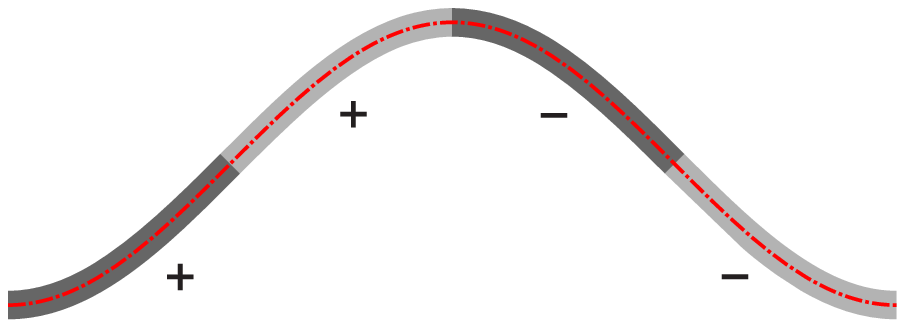}}
    \caption{These two DQPM structures have no kinks and allow studying a perturbative regime with only small 
deviations from the straight line. Therefore they will in practice allow balancing bend losses and nonlinear 
gain. (\textbf{Top:}) One period of this structure is composed of four circular arc segments that are each 
$L_c/2$ long. For large radii they will only attain small angles off the horizontal. (\textbf{Bottom:}) This 
structure is a cosine-shaped curve with a path length of $2L_c$. Because it has a continuously changing 
curvature it avoids the mode overlap losses that can happen in waveguides with a circular structure. However, 
at radii that are large with respect to the waveguide width the two structures become almost 
indistinguishable.}
    \label{fig:pdqpm}
\end{figure}

Figures~\ref{fig:2orderdqpm} and \ref{fig:pdqpm} depict second order and perturbative DQPM respectively.  These 
designs are intuitive because their subsections are in units or half units of $L_c$. This makes it relatively 
easy to keep track of the phase accrual between the pump and second harmonic fields.  The second order structure 
turns off the non-linearity whenever it's propagation axis lies along one of the crystallographic axis. Compared 
to first order DQPM, second order DQPM requires twice the distance to generate a similar intensity in the second 
harmonic.

The perturbative structure consists of two oppositely directed S-bends, each of length $L_c$, placed in series 
with each other. Each S-bend consists of two circular segments of length $L_c/2$. This produces a structure with 
smoothly varying angle $\alpha(s)$ and a curvature of constant magnitude but alternating 
sign.\textsuperscript{\footnotemark} Over the first S-bend, the transverse electric field direction stays in the 
quadrant where $\sin(2\alpha(s))$ is positive. Over the second S-bend, the electric field direction stays in the 
quadrant where $\sin(2\alpha(s))$ is negative. As in first order DQPM, this coincides with respective positive 
and negative contributions from the accumulated phase between the pump and second harmonic fields. Thus, the net 
result is always a positive contribution to the overall field amplitude of the second harmonic. The mild bending 
doesn't produce nearly the amount of polarization response in the medium and thus the conversion efficiency to 
the second harmonic is much less.

\footnotetext{A special case is a full circle, as it would occur in a ring resonator. In this case we would first 
have to solve for the ring resonance condition at the two frequencies and then check the coupling between the two 
modes, which is modulated sinusoidally around the ring with a $90^\circ$ period. Ignoring the ring resonator 
modes for a moment, we can integrate (\ref{eq:shgde}) around a ring with a circumference $2\pi r=N L_c$ that is 
an integer multiple of the nonlinear coherence length. We find that for $N\neq 4$ the field at $2\omega$ will 
always be identically zero after two loops. When $N=4$ a quasi-phase matching condition is achieved and the field 
will continue to grow. It may appear surprising at first that no higher-order QPM is possible, but if we realize 
that the ring yields a sinusoidal modulation of the effective nonlinearity, it is clear that there are no higher 
Fourier coefficients available for QPM. Therefore any ring-resonator QPM design seeks to use mode pairs with 
large $L_c$, in order to keep the ring radius reasonably big \cite{Yang07a, Kuo09}}

Depending on the precise modal structure of the waveguide a sudden change in curvature could be inducing more 
loss than an adiabatic one. Therefore we investigated cosine-shaped curves as alternatives to the segmented 
circular S-bend (see figure~\ref{fig:pdqpm}). In a cosine curve the curvature changes smoothly along the curve 
and thus mode overlap losses can be avoided. For large curvature radii and small lateral displacement the 
circular and cosine shaped S-bends become very similar both in shape and in performance.

\begin{figure}
    \centerline{\includegraphics[width=9cm]{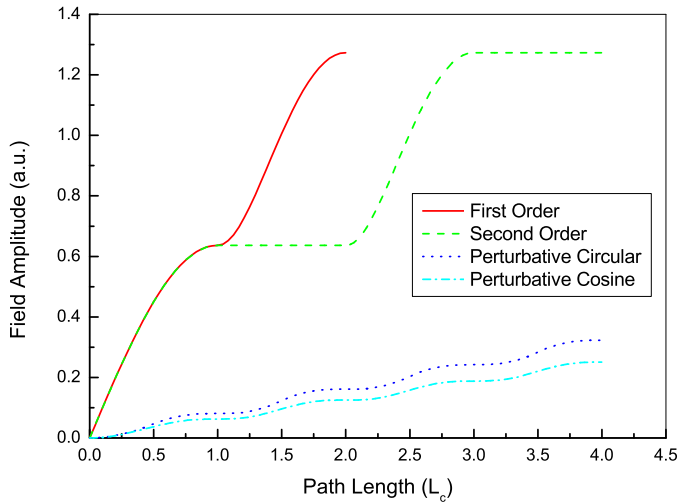}}
    \caption{Growth of the second harmonic field amplitude vs. the path length along the guide over two periods 
for first and second order DQPM, for the circular perturbative DQPM with $R=5 L_c$, and for the cosine-shaped 
perturbative DQPM with $R\geq5L_c$.}
    \label{fig:dqpm-shg}
\end{figure}

\begin{figure}
    \centerline{\includegraphics[width=8cm]{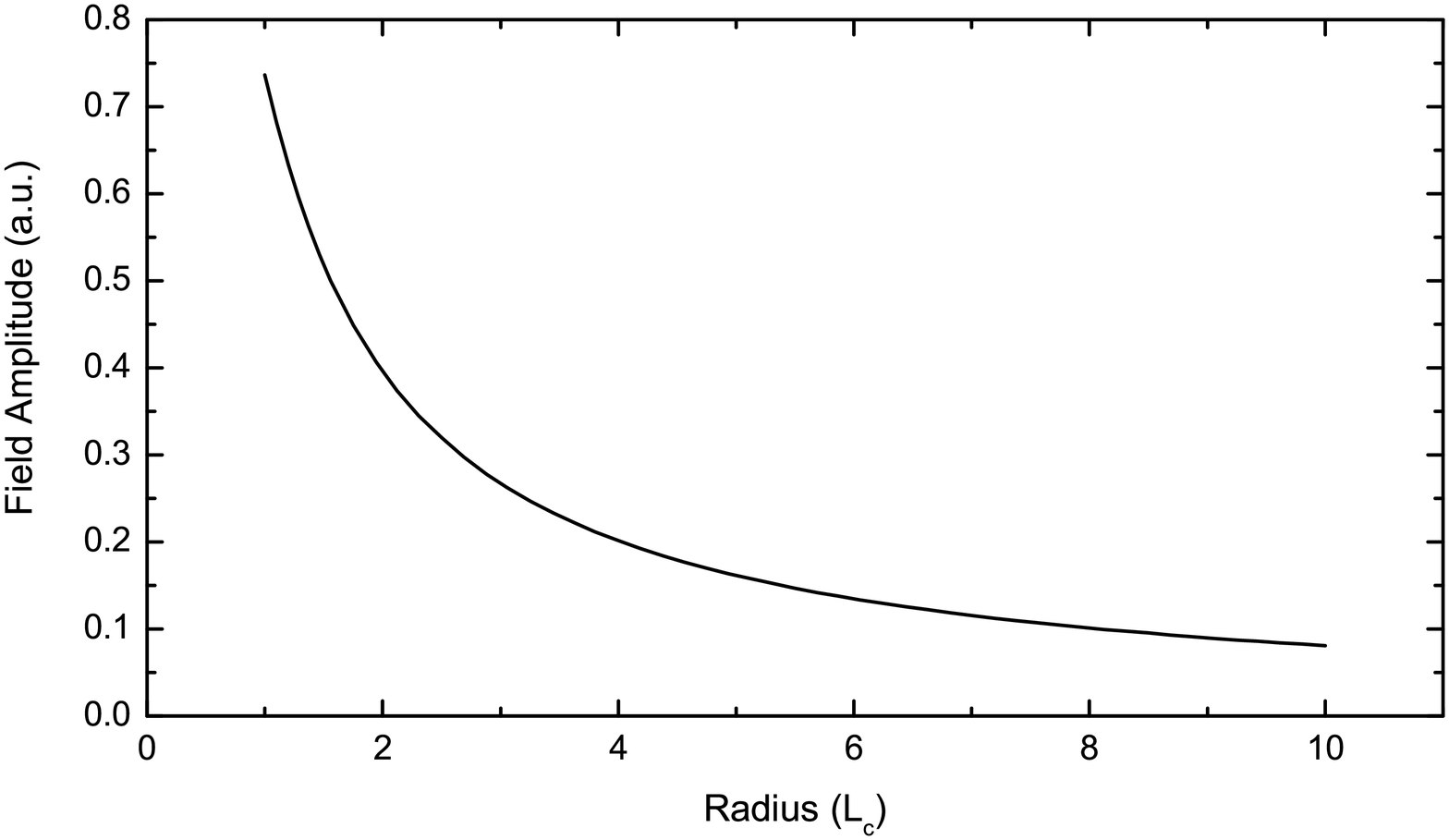}}
    \caption{SHG field amplitude after one period of the circular perturbative structure as a function of the 
curvature radius.}
    \label{fig:rdependence}
\end{figure}

Each design is scalable by virtue of the $2\pi$ relative phase between fundamental and second harmonic introduced 
by one unit cell and every period yields the same incremental amplitude in the undepleted pump approximation. 
This will lead to an overall quadratic growth of the second harmonic intensity. Figure~\ref{fig:dqpm-shg} plots 
the value of the integral (ignoring the constant prefactors) in (\ref{eq:shgde}) for two periods of each of the 
four structures. While (\ref{eq:shgde}) can be integrated analytically for all four structures with elliptic 
integrals for the cosine shaped curves, we used numeric integration to obtain these results, because of the 
greater flexibility. It is obvious that the more realistic, perturbative designs fall quite short of the maximum 
possible conversion efficiency for reasonable curvatures. With strong index contrast as, for example, in deeply 
etched ridge guides, these curvatures still seem feasible and it will remain to be shown by experiment how the 
nonlinear gain compares to the losses. Figure~\ref{fig:rdependence} shows the amplitude gain of one period of the 
circular structure as a function of the curvature radius $R$.

The ideal zig-zag of first order DQPM involves sudden changes in waveguide direction, and possibly violates some 
of the assumptions that underpin the basic analysis. Whether mediated by sharp bends or reflections, the design 
would almost certainly attenuate the field amplitudes after each successive direction change and possibly add a 
certain phase shift in the reflections.\textsuperscript{\footnotemark}

\footnotetext{For total internal reflection, this phase is precisely $\pi$, and for three waves, each wave would 
phase shift for a total change in the argument of the exponential of $\Delta{k}L_c$ + 3$\pi$ = 4$\pi$. This would 
mean that the phase has essentially already been reset by total internal reflection. An additional $\pi$ phase 
from the ensuing direction change in the guide would actually completely negate the SHG process This effect has 
been suggested \cite{Armstrong62} as a possible solution for QPM but the idea of potentially different coupling 
on the direction change was not mentioned.  For typical single mode waveguides in AlGaAs, L$_c$ is quite small, 
and to do effective phase matching in this manner would be quite difficult. One would need to either effect 
nearly 180$^\circ$ reflections or very shallow reflections in order to ignore the effects of DQPM.  Further it is 
unclear that the physics of total internal reflection would apply to standard waveguide modes.

In connection with strongly sloping sidewalls sharp bends can also induce facilitate the polarization rotations 
\cite{Dam96,Obayya02,Deng05}. With this one could conceivable also achieve some kind of QPM if the pump field 
could be made to rotate from in plane to out of plane with the correct periodicity.}

\begin{figure}
    \centerline{\includegraphics[width=9cm]{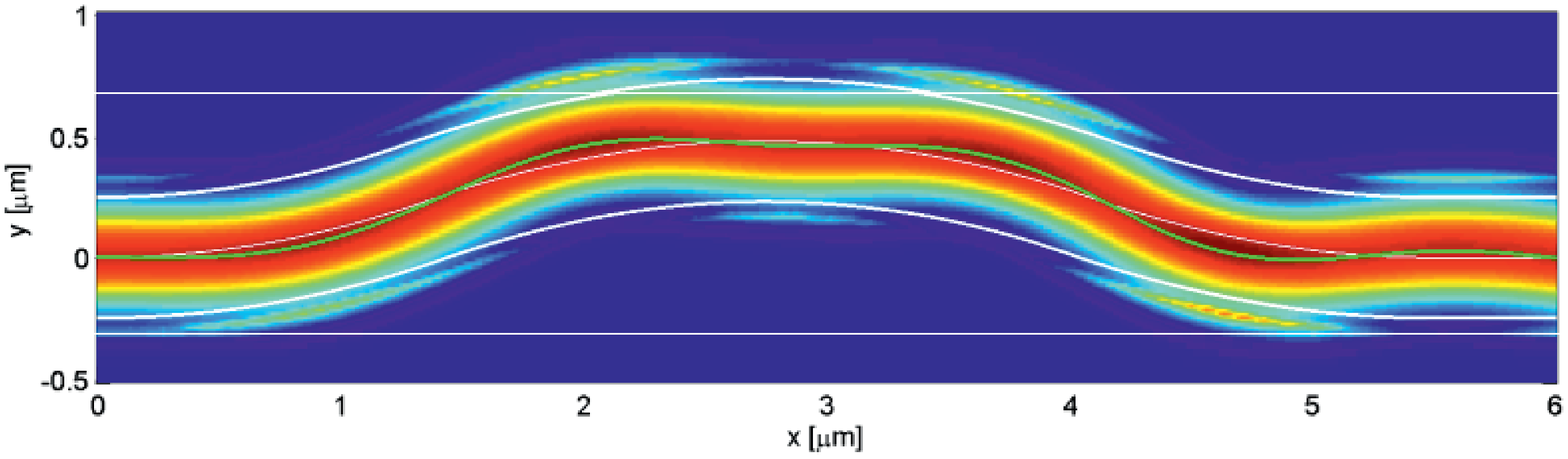}}
    \centerline{
        \includegraphics[height=3.5cm]{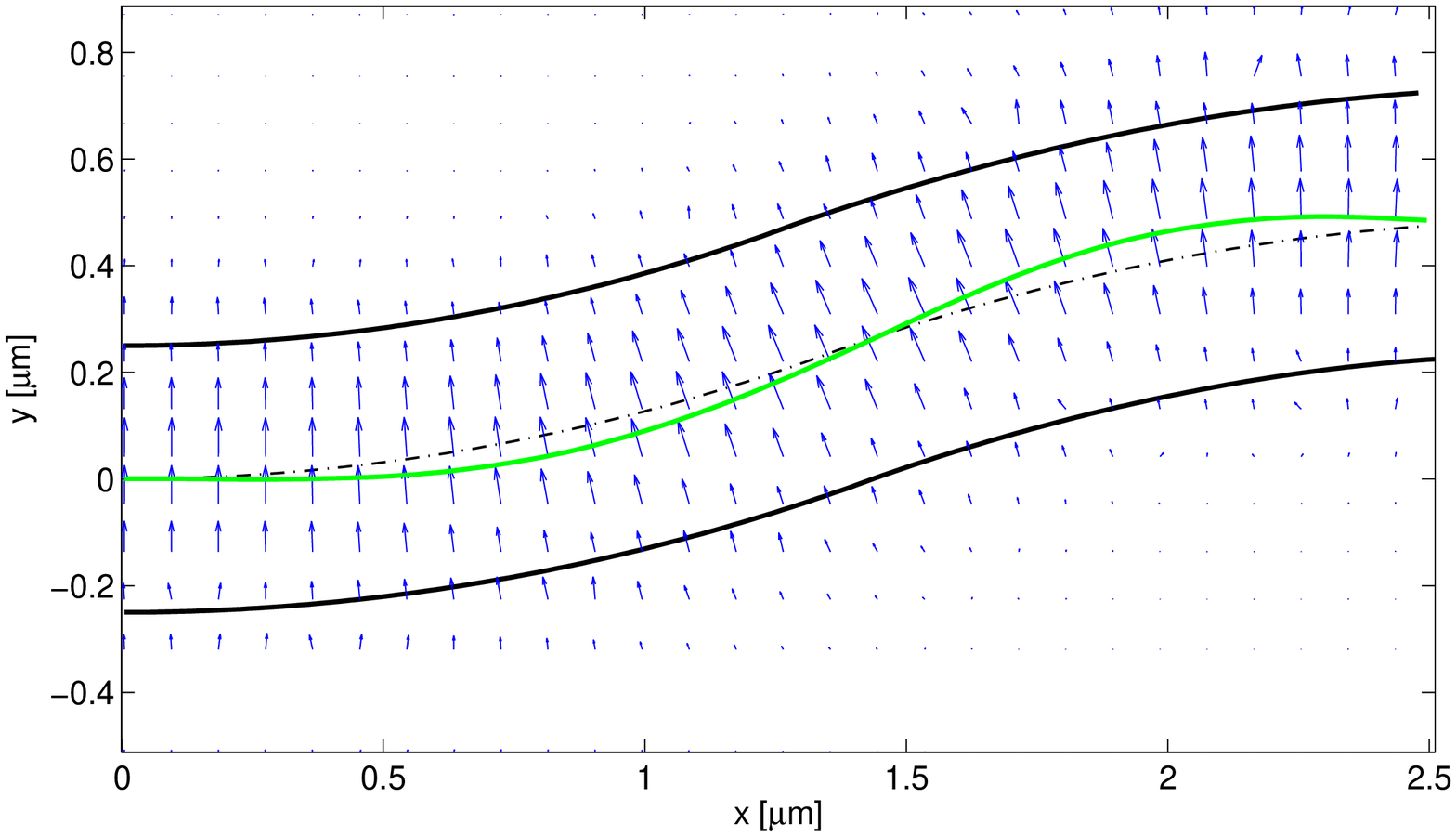} 
        \includegraphics[height=3.5cm]{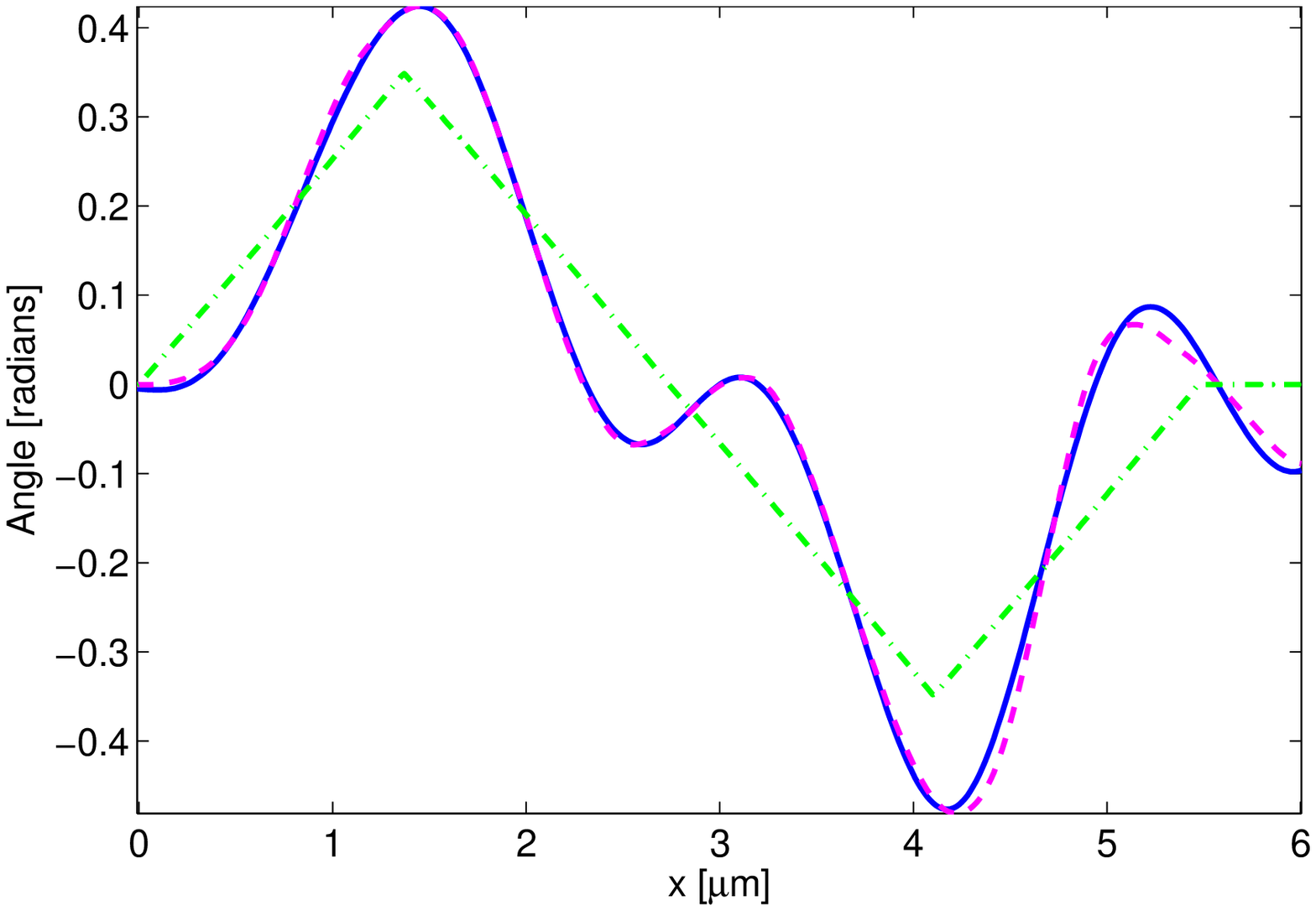}}
    \caption{Linear FDTD simulation of the circular perturbative DQPM structure. \textbf{Top:} False color plot 
of the intensity of TE (in-plane) polarized electric field at a vacuum wavelength of 1550~nm propagating 
through one period of the structure. The waveguide has refractive index 3.0, a width of 0.5~$\mu$m that and 
is surrounded by air. Here, the field is being injected in the fundamental waveguide mode from the left. It 
shows negligible loss but considerable field intensity outside the waveguide edges (thick white lines). 
\textbf{Bottom left:} Quiver plot of the local field polarization. The fact that this is always orthogonal to 
the local propagation direction can be visually verified along the path of the field maximum (thick green 
line). \textbf{Bottom right:} Local propagation direction (solid blue line) and the respective local 
polarization direction ($-90^\circ$, dashed magenta line) at the field maximum as a function of the 
horizontal coordinate along the waveguide. For comparison, the angle of the tangent to the waveguide 
centerline is shown with the dash-dotted green line.}
  \label{fig:fdtd}
\end{figure}

One of our simplifying assumptions is that the local TE polarization follows the normal to the waveguide 
centerline. It is clear that this cannot be true in the real situation. In order to get a feeling for the actual 
field path, we have performed two-dimensional Finite-Difference-Time-Domain (FDTD) simulations of the circular 
waveguide structure shown in the top part of figure~\ref{fig:pdqpm} using a commercial solver (Lumerical, FDTD 
Solutions 6.0). Figure~\ref{fig:fdtd} shows the power distribution, local polarization and the polarization angle 
along the field maximum. While we find that the field maximum takes a path that is quite different from the 
centerline, the polarization stays almost perfectly transverse, except near the edges of the guide, where it is 
comparably weak. This means that by integrating (\ref{eq:shgde}) along the actual field path one could design the 
ideal curvature and length to optimize the conversion efficiency. In practice, it would certainly be easier to 
use FDTD solvers that can treat the second order nonlinearity perturbatively while propagating the field and thus 
provide a complete simulation of the device. It is remarkable that the maximum curvature of the field path is 
quite a bit stronger than the waveguide's curvature. This means that with an adapted design we could expect a 
higher conversion efficiency than in our simple case.

The tight bends required by our proposed structures may seem very challenging, but a lot of theoretical and 
experimental research has been performed on this topic. Very tight bends \cite{Spiekman95,Jiang04}, including 
90$^\circ$ bends \cite{Espinola01} and more adiabatic bends \cite{Austin83,Deri87}, have been investigated. 
Important observations are: 1) the effective index decreases with decreasing bend radius; 2) wider guides exhibit 
more drastic modal changes in bends than do narrower guides; 3) Strong radial confinement is necessary to inhibit 
radiative loss in a bend; 4) high radial symmetry is necessary to reduce intermodal mixing. In essence coupling 
to radiative (non-guided) modes as well as intermodal coupling is inevitable in any waveguide structure that 
incorporates bending.

We have assumed that the mode profiles of the three interacting fields stay the same and that the ideal structure 
would enforce single mode behavior everywhere along it's length. In order to avoid intermodal mixing, excellent 
control over the spatial profile of the waveguide is desirable, which will typically result in very high aspect 
ratios. While this kind of structure can be fabricated, it requires state of the art manufacturing techniques 
that are extremely dependent on the proper chemistry during the fabrication process. Unfortunately, these 
processes also often have an unwanted side effect -- sidewall roughness. This will be the main source of loss due 
to the high index contrast and tight mode confinement for waveguides that have been etched through the core.

\begin{figure}
    \centerline{\includegraphics[width=6cm]{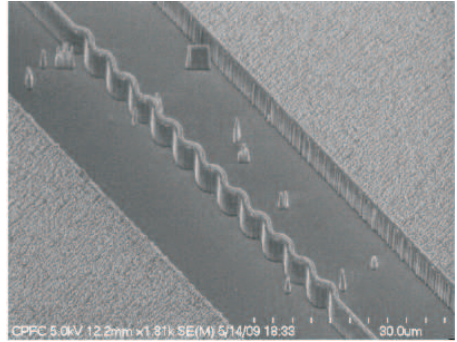}}
    \caption{A DQPM design featuring the perturbative scheme.  Here there are 13 unit cells repeated, and the 
perturbation is quite large so that bend losses are expected to be significant. We find that light is being 
guided through the structure but we only have very preliminary loss estimates.}
    \label{fig:bends-farview}
\end{figure}

Success of these designs will thus depend on how the conversion efficiency compares to the various loss 
mechanisms.  As there are in principle an infinite number of variations that might work, the actual design must 
incorporate fabrication methods as an input variable. A first fabrication try for a DQPM structure is shown in 
figure~\ref{fig:bends-farview}.

Modeling must now be performed to best predict design behavior based on the available fabrication capabilities.  
In the future, we hope to make considerations for such related issues like etching and regrowth processes and 
their effects on insertion and bend loss. Numerical simulations are underway to predict optimal mode matching 
between waveguides of different curvature and optimal aspect ratios.  Lastly, the basic theory may need some 
modification to allow for modal conversion along the waveguide structure. We believe that directional quasi-phase 
matching is an interesting and very flexible route towards exploiting second-order nonlinearities in many 
important semiconductor materials. These materials are notoriously hard to phase-match. Our concepts can be 
generalized to non-circular resonators, such as stadium resonators, in order to exploit higher-order 
phase-matching together with the increased field strength afforded by a cavity.

\ack

The authors wish to thank the Canadian Microsystems Corporation (CMC) for providing manufacturing resources and 
for their continued fabrication efforts, as well as J. Sipe for helpful discussions and insight into DQPM. This 
research has been funded in part by NSERC, CIFAR, CFI, CRC, ERA, OCE, and Quantum Works.

\section*{References}
\bibliographystyle{iopart-num}
\bibliography{dqpm}

\providecommand{\newblock}{}
\begin{thebibliography}{10}
\expandafter\ifx\csname url\endcsname\relax
  \def\url#1{{\tt #1}}\fi
\expandafter\ifx\csname urlprefix\endcsname\relax\def\urlprefix{URL }\fi
\providecommand{\eprint}[2][]{\url{#2}}

\bibitem{Kleinman66a}
Kleinman D~A, Ashkin A and Boyd G~D 1966 {\em Phys. Rev.\/} {\bf 145} 338--379

\bibitem{Kwiat95b}
Kwiat P~G, Mattle K, Weinfurter H, Zeilinger A, Sergienko A and Shih Y 1995
  {\em Phys. Rev. Lett.\/} {\bf 75} 4337--4341

\bibitem{Bocquillon09}
Bocquillon E, Couteau C, Razavi M, Laflamme R and Weihs G 2009 {\em Phys.
  Rev.~A\/} {\bf 79} 035801 (pages~4)
  \urlprefix\url{http://link.aps.org/abstract/PRA/v79/e035801}

\bibitem{Razavi09}
Razavi M, S\"ollner I, Bocquillon E, Couteau C, Laflamme R and Weihs G 2009
  {\em J. Phys.~B.\/} {\bf 42} 114013
  \urlprefix\url{http://stacks.iop.org/0953-4075/42/114013}

\bibitem{Bouwmeester97a}
Bouwmeester D, Pan J~W, Mattle K, Eibl M, Weinfurter H and Zeilinger A 1997
  {\em Nature\/} {\bf 390} 575--579

\bibitem{Erven08a}
Erven C, Couteau C, Laflamme R and Weihs G 2008 {\em Opt. Exp.\/} {\bf 16}
  16840--16853
  \urlprefix\url{http://www.opticsexpress.org/abstract.cfm?URI=oe-16-21-16840}

\bibitem{Pan03b}
Pan J~W, Gasparoni S, Ursin R, Weihs G and Zeilinger A 2003 {\em Nature\/} {\bf
  423} 417--422

\bibitem{Scaccabarozzi06a}
Scaccabarozzi L, Fejer M~M, Huo Y, Fan S, Yu X and Harris J~S 2006 {\em Opt.
  Lett.\/} {\bf 31} 3626--3528

\bibitem{Leo99}
Leo G, Berger V, OwYang C and Nagle J 1999 {\em J. Opt. Soc. Am. B\/} {\bf 16}
  1597--1602
  \urlprefix\url{http://josab.osa.org/abstract.cfm?URI=josab-16-9-1597}

\bibitem{Helmy06a}
Helmy A~S 2006 {\em Opt. Exp.\/} {\bf 14} 1243--1252

\bibitem{Armstrong62}
Armstrong J~A, Bloembergen N, Ducuing J and Pershan P~S 1962 {\em Phys. Rev.\/}
  {\bf 127} 1918--1939

\bibitem{Weihs06}
Weihs G 2006 {\em Int. J. Mod. Phys. B\/} {\bf 20} 1543--1550
  (\textit{Preprint} \eprint{http://arxiv.org/abs/quant-ph/0508179})

\bibitem{Boyd08}
Boyd R~W 2008 {\em Nonlinear Optics\/} 3rd ed (Burlington, MA: Academic Press)

\bibitem{Haidar04}
Ha\"{i}dar R, Forget N, Kupecek P and Rosencher E 2004 {\em J. Opt. Soc.
  Am.~B\/} {\bf 21} 1522--1534
  \urlprefix\url{http://josab.osa.org/abstract.cfm?URI=josab-21-8-1522}

\bibitem{Dumeige06a}
Dumeige Y and Feron P 2006 {\em Phys. Rev.~A\/} {\bf 74} 063804 (pages~7)
  \urlprefix\url{http://link.aps.org/abstract/PRA/v74/e063804}

\bibitem{Yang07a}
Yang Z, Chak P, Bristow A~D, van Driel H~M, Iyer R, Aitchison J~S, Smirl A~L
  and Sipe J~E 2007 {\em Opt. Lett.\/} {\bf 32} 826--828
  \urlprefix\url{http://www.opticsinfobase.org/abstract.cfm?URI=ol-32-7-826}

\bibitem{Kuo09}
Kuo P~S, Fang W and Solomon G~S 2009 {\em Opt. Lett.\/} {\bf 34} 3580--3582
  \urlprefix\url{http://ol.osa.org/abstract.cfm?URI=ol-34-22-3580}

\bibitem{Kim06}
Kim S, Jiang J and Nordin G~P 2006 {\em Optical Engineering\/} {\bf 45} 054602

\bibitem{Dam96}
van Dam C, Spiekman L, van Ham F, Groen F, van~der Tol J, Moerman I, Pascher W,
  Hamacher M, Heidrich H, Weinert C and Smit M 1996 {\em IEEE Photonics
  Technology Letters\/} {\bf 8} 1346--1348 ISSN 1041-1135

\bibitem{Obayya02}
Obayya S, Rahman B, Grattan K and El-Mikati H 2002 {\em IEE Proc.
  Optoelectronics\/} {\bf 149} 75--80
  \urlprefix\url{http://link.aip.org/link/?IPO/149/75/1}

\bibitem{Deng05}
Deng H, Yevick D and Chaudhuri S 2005 {\em IEEE Photonics Technology Letters\/}
  {\bf 17} 2113--2115 ISSN 1041-1135

\bibitem{Spiekman95}
Spiekman L, Oei Y, Metaal E, Groen F, Demeester P and Smit M 1995 {\em IEE
  Proc. Optoelectronics\/} {\bf 142} 61--65 ISSN 1350-2433

\bibitem{Jiang04}
Jiang A, Shi S, Jin G and Prather D 2004 {\em Opt. Express\/} {\bf 12} 633--643
  \urlprefix\url{http://www.opticsexpress.org/abstract.cfm?URI=oe-12-4-633}

\bibitem{Espinola01}
Espinola R, Ahmad R, Pizzuto F, Steel M and Osgood R 2001 {\em Opt. Express\/}
  {\bf 8} 517--528
  \urlprefix\url{http://www.opticsexpress.org/abstract.cfm?URI=oe-8-9-517}

\bibitem{Austin83}
{Austin} M~W and {Flavin} P~G 1983 {\em Journal of Lightwave Technology\/} {\bf
  1} 236--240

\bibitem{Deri87}
Deri R, Kapon E and Schiavone L 1987 {\em Electron. Lett.\/} {\bf 23} 845--847

\end{thebibliography}

\end{document}